%% file: sosp25-ams.tex
\documentclass[sigplan,10pt,nonacm]{acmart}
\renewcommand\footnotetextcopyrightpermission[1]{}
\AtBeginDocument{%
  }
\usepackage{multirow}
\usepackage{algorithm}
\usepackage{algorithmic}
\usepackage{soul}

\newcommand{\FEH}[1]{{#1}}
\newcommand{\FEHSOSP}[1]{{#1}}

\newcommand{\REFINE}[1]{{#1}}

\newcommand{\myparagraph}[1]{\noindent\textbf{\emph{#1}}}

\renewcommand{\shortauthors}{}

\newcommand{\sys}{{\texttt{AMS}}}
\newcommand{\sysintr}{action interrupt}

\newcommand{\sysactL}{Action}
\newcommand{\sysctx}{action context}
\newcommand{\sysctxL}{Action Context}
\newcommand{\sysexcept}{action exception}
\newcommand{\sysexceptL}{Action Exception}
\newcommand{\sysreplay}{action replay}
\newcommand{\sysreplayL}{Action Replay}
\newcommand{\sysslice}{action slice}

\begin{document}

\title[Leveraging OS-Level Primitives for Robotic Action Management]{Leveraging OS-Level Primitives for Robotic Action Management}



\author{Wenxin Zheng}
\affiliation{%
  \institution{Shanghai Jiao Tong University}
  \city{Shanghai}
  \country{China}}
\email{wxzheng98@gmail.com}

\author{Boyang Li}
\affiliation{%
  \institution{Southern University of Science and Technology}
  \city{Shenzhen}
  \country{China}}
\email{12111014@mail.sustech.edu.cn}

\author{Bin Xu}
\affiliation{%
  \institution{Shanghai Jiao Tong University}
  \city{Shanghai}
  \country{China}}
\email{levi.xu.bin@gmail.com}

\author{Erhu Feng}
\affiliation{%
  \institution{Shanghai Jiao Tong University}
  \city{Shanghai}
  \country{China}}
\email{fengerhu1@sjtu.edu.cn}

\author{Jinyu Gu}
\affiliation{%
  \institution{Shanghai Jiao Tong University}
  \city{Shanghai}
  \country{China}}
\email{gujinyu@sjtu.edu.cn}

\author{Haibo Chen}
\affiliation{%
  \institution{Shanghai Jiao Tong University}
  \city{Shanghai}
  \country{China}}
\email{haibochen@sjtu.edu.cn}







\renewcommand{\shortauthors}{Wenxin Zheng, Boyang Li, Bin Xu, Erhu Feng, Jinyu Gu, Haibo Chen}

\input{src/0-abs}

\settopmatter{printfolios=true}
\maketitle
\input{src/1-intro}
\input{src/2-motivation}

\input{src/3-design}

\input{src/4-impl}

\input{src/6-eval}

\input{src/7-related-conclu}
\bibliographystyle{ACM-Reference-Format}
\bibliography{sosp25-ams}










\end{document}

%% file: src/0-abs.tex
\begin{abstract}
    \FEH{End-to-end imitation learning frameworks (e.g., VLA) are increasingly prominent in robotics, 
    as they enable rapid task transfer by learning directly from perception to control, 
    eliminating the need for complex hand-crafted features.
    However, even when employing SOTA VLA-based models, they still exhibit limited generalization capabilities and suboptimal action efficiency,
    due to the constraints imposed by insufficient robotic training datasets.
    In addition to addressing this problem using model-based approaches, 
    we observe that robotic action slices, which consist of contiguous action steps,
    exhibit strong analogies to the time slices of threads in traditional operating systems. 
    This insight presents a novel opportunity to tackle the problem at the system level.}

    \FEH{In this paper, we propose \sys{}, a robot action management system enhanced with OS-level primitives like \textit{exception}, \textit{context switch} and \textit{record-and-replay},
    that improves both execution efficiency and success rates of robotic tasks. 
    \sys{} first introduces \sysexcept{}, 
    which facilitates the immediate interruption of robotic actions to prevent error propagation. 
    Secondly, \sys{} proposes \sysctx{}, which eliminates redundant computations for VLA-based models, 
    thereby accelerating execution efficiency in robotic actions. 
    Finally, \sys{} leverages \sysreplay{} to facilitate repetitive or similar robotic tasks without the need for re-training efforts.
    We implement \sys{} in both an emulated environment and on a real robot platform. 
    The evaluation results demonstrate that \sys{} significantly enhances the model's generalization ability and action efficiency, 
    achieving task success rate improvements ranging from 7$\times$ to 24$\times$ and saving end-to-end execution time ranging from 29\% to 74\% compared to existing robotic system without \sys{} support.  
    }

\end{abstract}

%% file: src/1-intro.tex
\section{Introduction}

End-to-end policy learning framework are gaining increasing attention in the field of robotics. 
Researchers believe that this robotic computing paradigm with imitation learning~\cite{rt1:brohan2022rt,xi2025rise,driess2023palm,fu2024mobile} is the future for rapid task transfer 
because these models can learn directly from perception to control, 
eliminating the complex hand-crafted feature extraction and modular design of traditional methods. 
Through training on large-scale datasets, end-to-end models are expected to autonomously learn to adapt in complex and changing environments, 
thereby exhibiting greater flexibility and efficiency in perception, decision-making, and action execution. 

\begin{figure}[!t]
    \centering
    \includegraphics[width=\columnwidth]{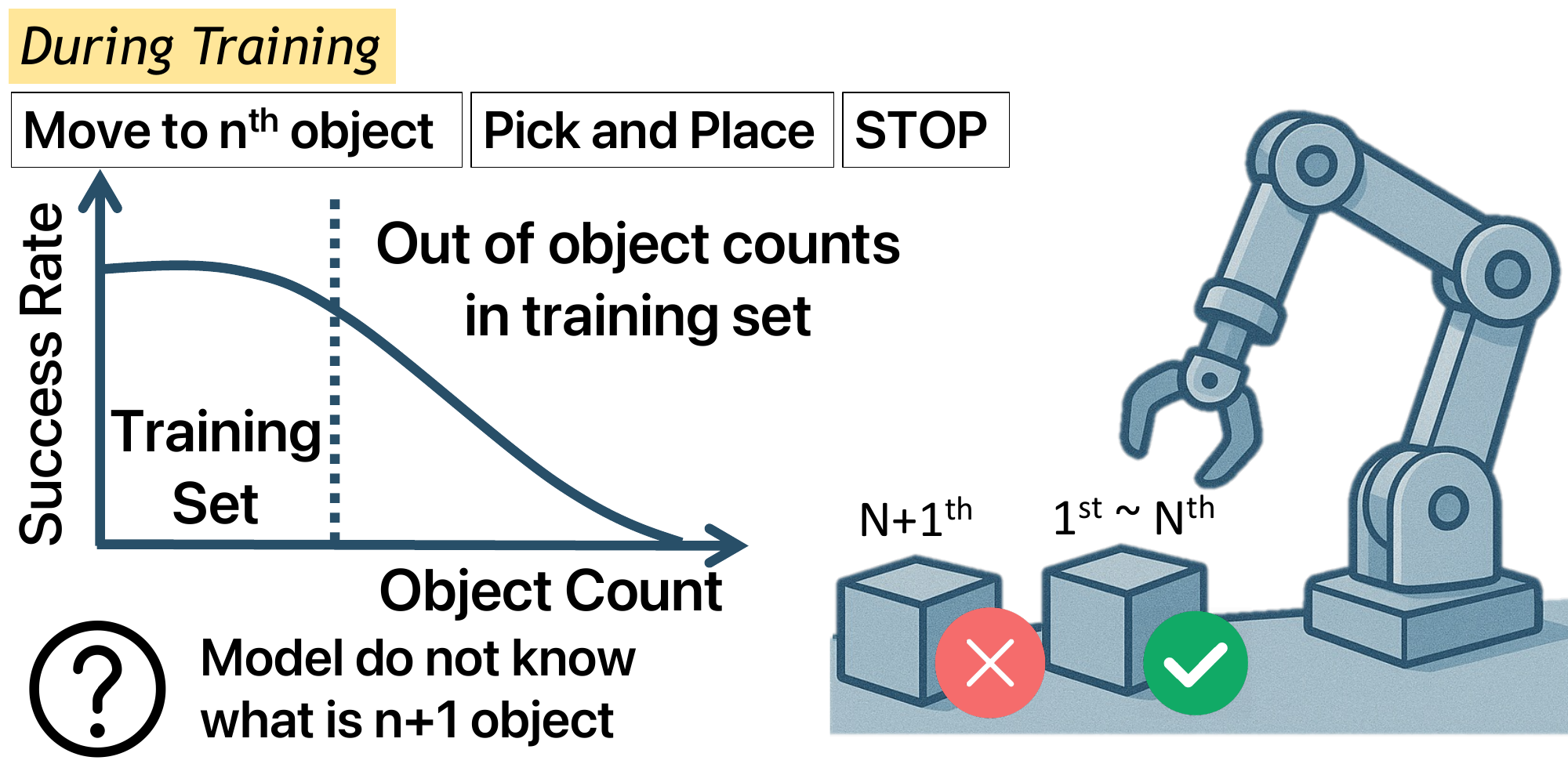}
    \caption{\textbf{A failure case of the robot's pick-and-place operation. } 
    When the number of objects to be picked is not represented in the training set, 
    the action's success rate results in a significant decrease.}
    \label{fig:intro:false:stop}
\end{figure}

\FEHSOSP{However, current robotic models still exhibit limitations in their generalization capabilities~\cite{ze2024generalizable,wang2024rise} and action performance. 
For instance, in a typical robotic task such as pick and place, 
as illustrated in Figure~\ref{fig:intro:false:stop}, 
the success rate declines dramatically once the number of objects exceeds the maximum encountered in the training set, 
even when utilizing SOTA robotic models~\cite{chen2023polarnet,goyal2023rvt,guhur2023instruction,riegler2017octnet,rt1:brohan2022rt}. 
This is primarily due to the limited availability of training data for real robotic platforms, 
which often leads to model overfitting and consequently results in insufficient generalization capabilities, 
even for tasks that share similar characteristics.
Furthermore, considering the action efficiency, 
the robot's actions are typically slower than those of human beings.
This is primarily due to the limited actions per second (APS) during model inference, 
as well as the interference of generated actions across different objects, which  results in non-optimal trajectory paths.
}

To address this limitation, the robotics research community focuses on improving the model's generalization ability from an algorithmic perspective. 
This includes introducing more environmental information~\cite{chen2023polarnet,goyal2023rvt,guhur2023instruction,riegler2017octnet}, using LLM ability~\cite{ha2023scaling}, optimizing model parameters~\cite{park2024quantization,lin2025awq,yue2024deer,zhao2024dynamic,zhang2024sparsevlm,james2022coarse}, 
and designing entirely new network architectures to enhance the model's adaptability in unseen situations~\cite{wang2024rise,gervet2023act3d,chi2023diffusion,riegler2017octnet,rt1:brohan2022rt}. 
\FEHSOSP{Unfortunately, these methods still face limitations in practical applications, 
such as the scarcity of training data for real robotic platforms and the retraining overhead associated with different tasks and robotic systems.}

\FEH{After a thorough analysis of the current robotic workflow, 
we identify a new opportunity to enhance robotic robustness and efficiency from the OS level. 
Although the tasks executed in robotic systems differ from those of traditional applications, 
we can still discern similarities between these two workloads and leverage the primitives offered by classical operating systems.
For example, to ensure the coherence of robotic actions, 
the current robotic model outputs a fixed number of action steps during each inference. 
We refer to this sequence of action steps as an ``action slice''. 
The action slice is similar to the time slice allocated for application threads in traditional operating systems, 
yet it lacks certain system capabilities.
For example, in the current robotic workload, 
the action slice cannot be interrupted even when a fault situation arises, 
leading to a significant degradation in the success rate of robotic tasks.
Furthermore, after executing one action slice, all context (e.g., intermediate result for robotic models) are discarded, 
and the robot generates the next action slice from an initialized state, 
which limits the performance of robotic actions.}

\FEH{\FEHSOSP{Inspired by traditional operating system primitives such as context switching, exception handling, and record-and-replay mechanisms,}
\sys{} introduces \textbf{\sysctx{}}, \textbf{\sysexcept{}}, and \textbf{\sysreplay{}} for robotic fault detection, 
context saving and restoration across different action slices, 
and replaying robotic actions within similar environments. 
With these OS primitives, the efficiency and success rate of robotic actions are significantly enhanced without the necessity of retraining the model.
Specifically, \sys{} proposes three key technical designs:
}

\FEH{Firstly, \sys{} stores and manages all \sysctx{} within a context pool. 
The \sysctx{} encompasses all intermediate inference states, 
such as KV caches, latent vectors, and output embeddings for robotic models. 
Unlike traditional robotic systems, where each action is generated from a clean initial state, \sys{} reuses the context generated from the previous round of inference. 
This approach significantly reduces redundant computations, accelerates model inference, and decreases the number of actions steps.
To mitigate storage overhead, \sys{} employs hierarchical and differential storage strategies, 
retaining only the essential and distinctive components closer to the computing resources while evicting/offloading redundant or insignificant \sysctx{}.}

\FEH{Secondly, \sys{} implements GPU-Inference with CPU-Action-Fault-Detection to support \sysexcept{}.
It contains multiple rule-based functions that run on the CPU side to detect any hardware/software-defined robotic exceptions, 
and interrupt the execution of robotic action immediately.
By preventing the propagation of faults, robots are more likely to fix erroneous actions with the dedicated exception handler.}

\FEH{Thirdly, \FEHSOSP{to enhance the generalization capability in a typical robotic scenario involving long-horizon tasks that include multiple repetitive jobs, 
\sys{} proposes \sysreplay{}, which leverages previous successful contexts to replay robot actions in similar environments.}
However, unlike the fixed record-and-replay processes in traditional operating systems, 
in robotic scenarios, \sys{} can dynamically regenerate the replay actions based on environmental changes.
}

\FEH{Our prototype of \sys{} is implemented as an OS service with several extensive modules, 
including context manager and exception tables, etc.
We effectively manage context information by creating context pools on both the GPU and CPU, 
along with the hashed action index, virtual actions, and an automatic action eviction mechanism.
Regarding model selection, we utilize a SOTA VLA-based robotic model $\pi_0$~\cite{black2024pi0visionlanguageactionflowmodel} without requiring retraining efforts for tasks. 
We believe that the \sys{} approach is sufficiently general to be adaptable to various models and robotic configurations.
}

We evaluate the \sys{} on a real robot platform and an emulated environment.
Our test results show that \sys{} can effectively improve the model's generalization ability, 
and reduce the time needed for the model to perform whole tasks, both achieving better and faster performance over time. 
In real robot tests, the success rate of \FEHSOSP{long-horizon task} inference after activating \sys{} is 7$\times$ to 24$\times$ higher than that of direct model inference. 
Regarding the execution time of the whole action, 
after activating \sys{}, the number of steps required by the robot decreased by 29\% to 74.4\% compared to direct model inference. 
Compared to the time required for the first inference, 
\sys{} can reduce the time of subsequent inferences by 5.7\% to 20\%.

%% file: src/2-motivation.tex
\section{Background and Motivation}

\subsection{Large Models in Robotics}
\begin{figure}[!ht]
    \centering
    \includegraphics[width=\columnwidth]{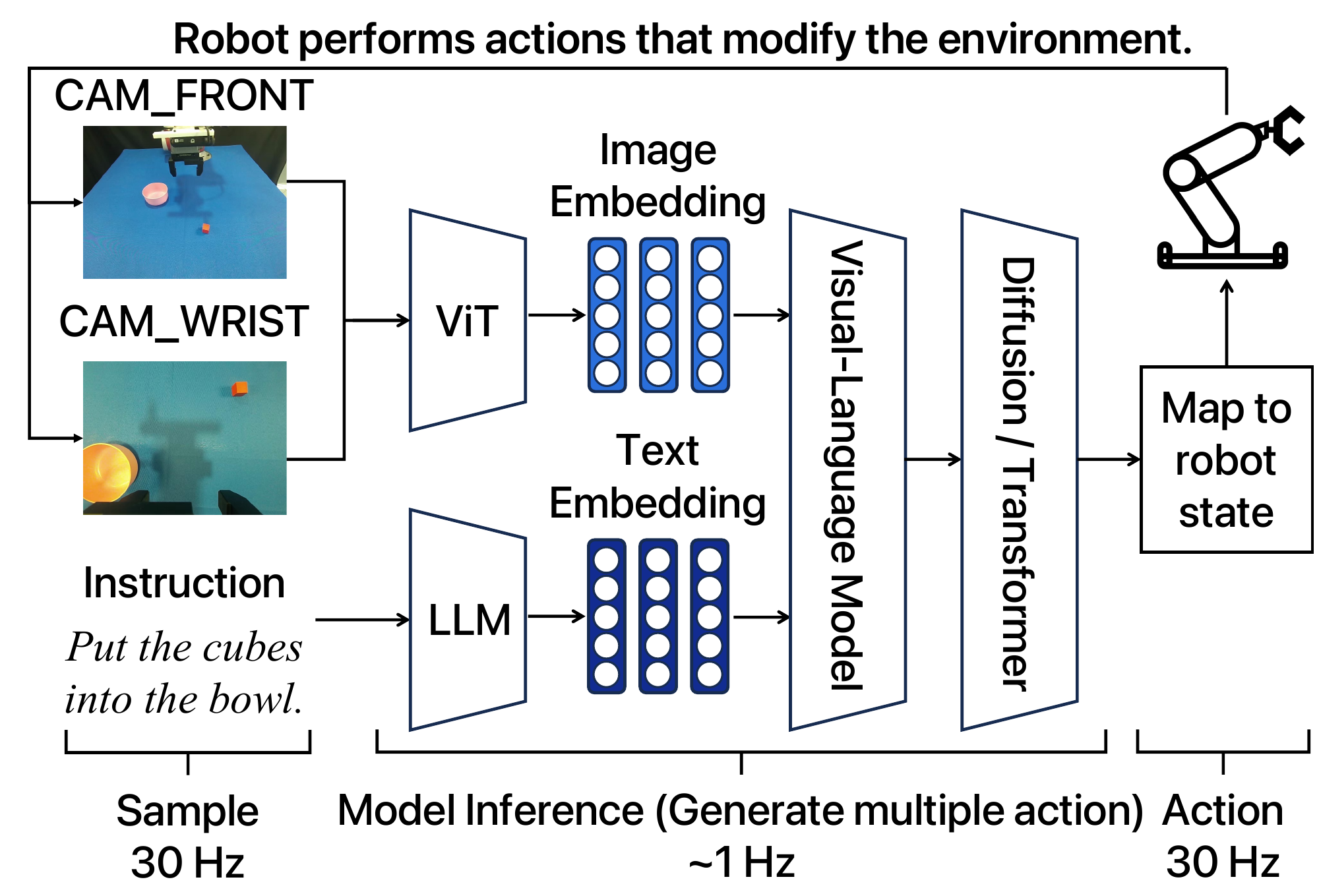}
    \caption{\textbf{State-of-the-art VLA model structure. }\FEHSOSP{It usually combines three models: visual encoders, large language model, and action decoders (e.g., diffusion model).}}
    \label{fig:motiv:vla}
\end{figure}

As the size and capabilities of models continue to scale up, 
end-to-end learning frameworks of large models becomes more and more important in robotic manipulations, 
\FEH{especially for current Vision-Language-Action (VLA) Models~\cite{black2024pi0visionlanguageactionflowmodel,zhang2025up,zhao2025cot,kim2024openvla,guo2025improving,adilkhanov2025survey,zhen20243d,chen2024meshxl,li2024llara,xu2024humanvla,wen2025tinyvla}.} 
These models generally consist of three key components: natural language processing, environment perception, and action planning.
Typically, LLMs are used for natural language processing, multi-modal large models are used as environmental perception components and diffusion model is used for real action planning.
Different models have different inference frequency, as shown in~\autoref{fig:motiv:vla}.

The natural language processing component is responsible for interpreting the user's natural language commands and converting them to text vectors. 
The environment perception component gathers information about the surroundings.
It collects 2D images or 3D point cloud data with depth information by the robot's sensors, such as cameras and LiDAR. 
These 2D or 3D images are then transformed into image vectors using Transformer-based models like Vision-in-Transformer (ViT)~\FEH{\cite{radford2021learningtransferablevisualmodels,dosovitskiy2021imageworth16x16words}}.
Building on this, the action planning component utilizes text and image vectors to generate a series of motion commands suitable for the robot and an optional stop flag. 
The action planning components are mainly diffusion models~\cite{carvalho2024motion,wen2024diffusion,hou2025dita,pan2024vision,black2024pi0visionlanguageactionflowmodel,janner2022planning,chi2023diffusion} or auto-regressive models\FEH{~\cite{kim2024openvla,zhang2025autoregressive,jiang2024survey,xie2024towards,choimixture,george2023one,zhao2023learning,park2024hierarchical}}. 

Different robotic arms have different degrees of freedom, so the models cannot be directly shared between different robotic arms.
The model parameters from the large-scale pre-trained model needs to be fine-tuned on each robotic arm individually to adapt it to each robotic arm.

\subsection{Current Limitation in VLA-based Robotic Models}

\subsubsection{Generalization Problems}
Currently, the generalization ability of robotic models relies mainly on fitting limited data during training. 
They adjust their parameters by learning patterns and features from this training data. 
However, unlike large language models, robots operate in complex and ever-changing environments. 
Gathering large and diverse datasets of robot actions is very difficult and requires huge human efforts, which limits the diversity and scale of their training data. 
This lack of data hinders the development of robotic models, preventing them from showing new abilities in unfamiliar situations like large language models can\FEH{~\cite{firoozi2023foundation}}.

\begin{figure}[!ht]
    \centering
    \includegraphics[width=\columnwidth]{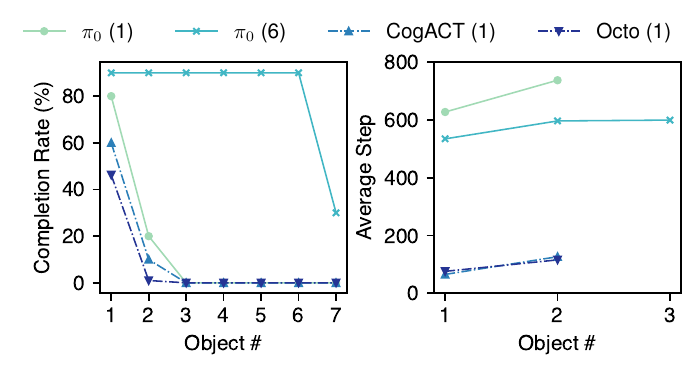}
    \caption{\textbf{Degrading of different models.} The number in the brackets means the maximum repetition counts in the training set.}
    \label{fig:motiv:degrading}
\end{figure}

Due to these training data limitations, robotic models perform poorly when asked to do actions beyond what they've seen before. 
For example, \FEHSOSP{a long-horizon task involving multiple repetitive or similar jobs is a typical scenario in robotics, such as the pick-and-place operation of multiple objects.}
The \FEH{$\pi_0$} models~\cite{black2024pi0visionlanguageactionflowmodel} typically don't exceed the maximum number of repetitions they encountered during training. 
We tested different models with various repetition counts in their training data. 
The results in~\autoref{fig:motiv:degrading} showed that models had high success rates up to the maximum repetition seen during training, but their accuracy dropped significantly to zero beyond that point. 

\begin{figure}[!ht]
    \centering
    \includegraphics[width=\columnwidth]{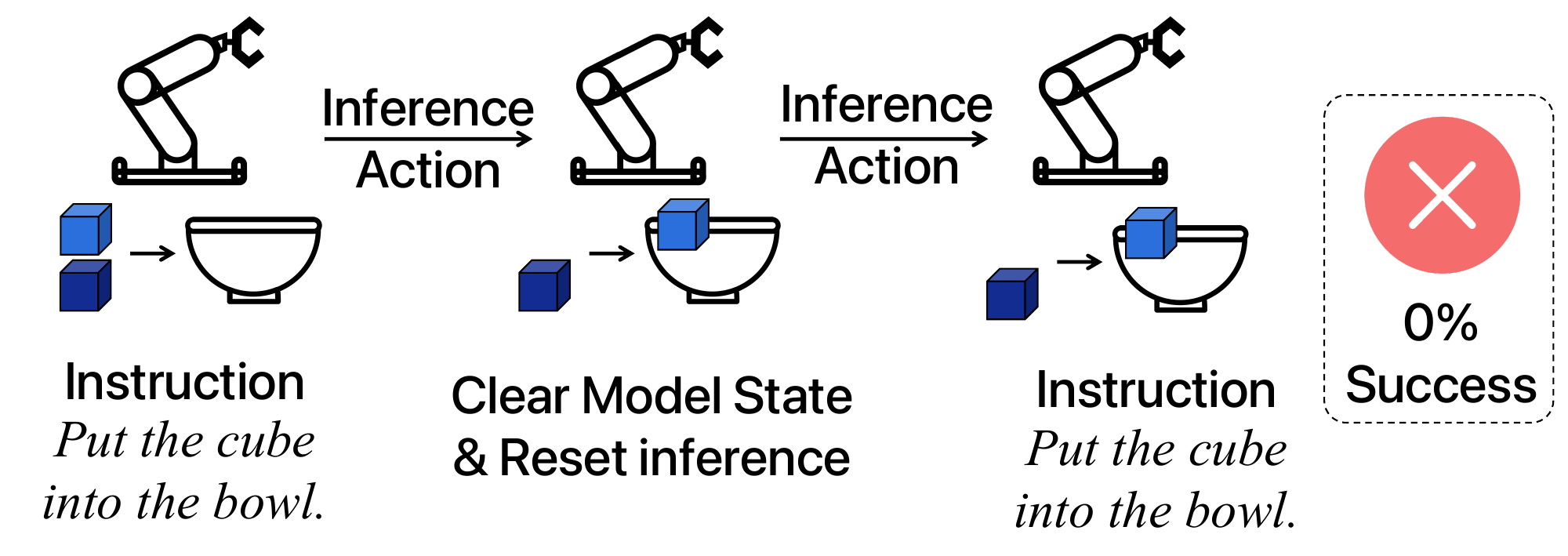}
    \caption{\textbf{Strawman design of resetting the model inference.}}
    \label{fig:motiv:reset-degrading}
\end{figure}

\FEH{\textbf{Key Challenges:} To enhance the model's generalization ability, 
\FEHSOSP{our work mainly focuses on classical robotic scenarios, specifically long-horizon tasks that involve multiple repetitive jobs.}
We first adopt a straw-man design: resetting the model's internal state and regenerating robotic actions. 
\autoref{fig:motiv:reset-degrading} illustrates a classical robotic task of ``picking up all small blocks and placing them into a bow'' (``pick and place''). 
In its training set, the robot learns only to place one block into the bowl and has never encountered a situation involving two blocks. 
Employing the straw-man method fails to yield even a single successful completion. 
This failure occurs because the environmental state after placing one block leads the model to mistakenly believe that it has completed the entire task, 
thereby preventing the robot from executing subsequent actions. 
Consequently, the key challenge for model generalization in these typical robotic tasks is 
\FEHSOSP{\textbf{how to enable the model to perform repetitive and similar jobs within a long-horizon task without the need for retraining efforts.}}
}


\subsubsection{Efficiency Problems}
\REFINE{Efficiency is also an important part of a model's performance. 
The end-to-end time relates to actions per second (APS) and the total number of action steps. 
\FEHSOSP{The APS is influenced by two factors: the speed of model inference and the performance of the robotic hardware. 
The minimum value among these two factors determines the actual APS value observed in the test.}
As for total number of steps, it depends on the results of each inference. 
Humans learn from repeating tasks, dropping unnecessary actions and correcting mistakes to become more skilled over time. 
However, robots cannot learn from past tasks.
For instance, when the model has already kicked a ball and its position is very close to the shortest path, 
the model still cannot find that optimal path to kick the ball. 
They rely heavily on the training data, which can lead to performing tasks slower over time, as shown in~\autoref{fig:motiv:degrading}.}

\textbf{Key Challenges: } 
Previous work~\cite{xu2025vla} has attempted to enhance the efficiency of robot execution by accelerating model inference. 
However, in real robotic scenarios, hardware performance often constitutes the primary bottleneck, 
resulting in only minimal improvements from optimizing model inference speed. 
Consequently, the primary challenge for enhancing robotic efficiency lies in 
\FEHSOSP{how to enable the model to identify the optimal action path, thereby reducing the total number of action steps.}

%% file: src/3-design.tex
\section{\sys{} Designs}

\begin{figure}
    \centering
    \includegraphics[width=\linewidth]{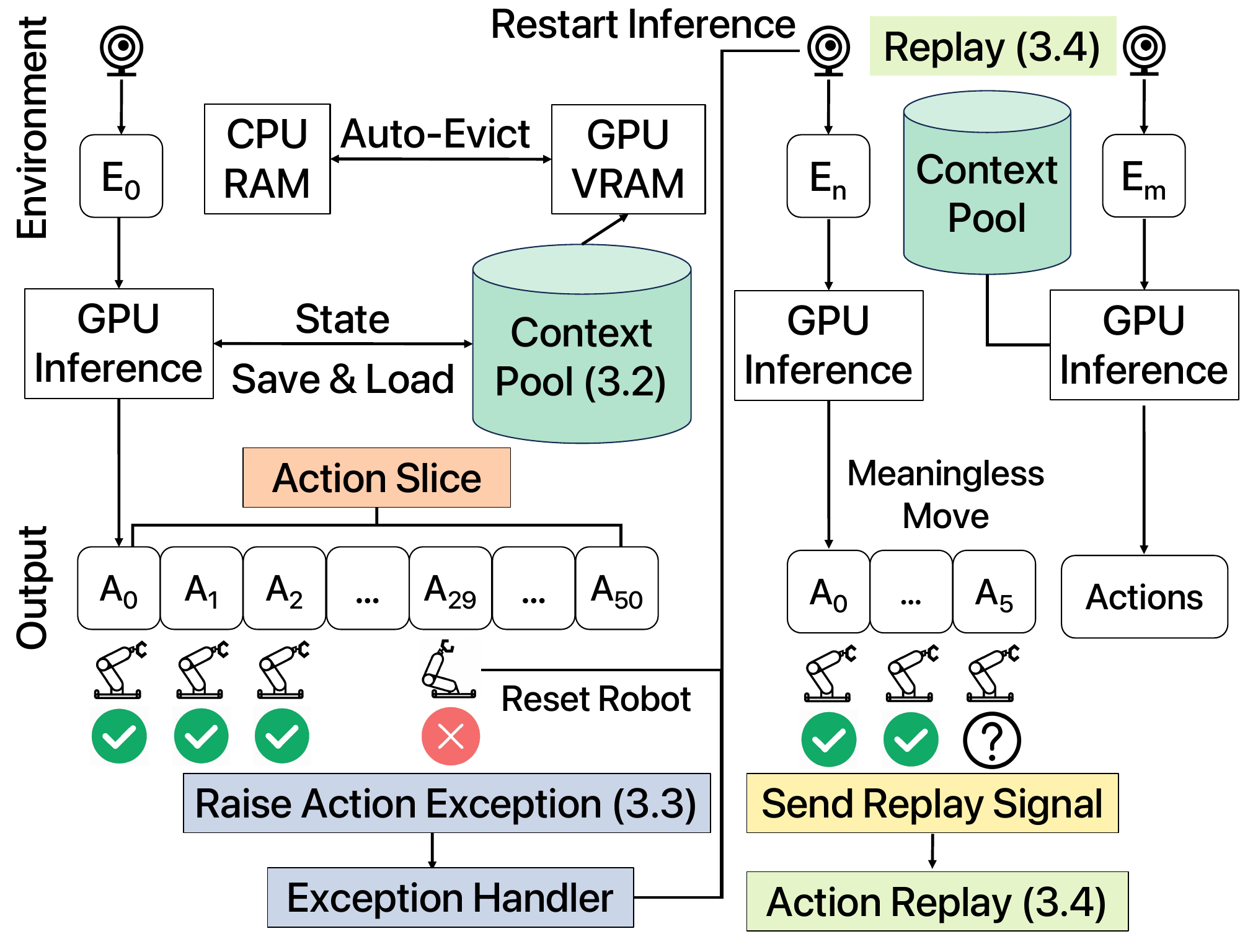}
    \caption{\textbf{The overall design of \sys{} enabled inference process for robotic models.}}
    \label{fig:concepts:full-new}
\end{figure}

\subsection{Key Insight}
\FEH{To enhance the efficiency and success rate of robotic actions, 
the core of \sys{} is to \textbf{migrate OS-level primitives into robotic action management}. 
Similar to the time slice abstraction of threads in traditional operating systems, 
the action slice in the robotic system, which comprises several contiguous action steps, 
serves as the minimal and atomic execution unit. 
However, in traditional operating systems, when a program encounters an exception or fault during execution, 
the kernel interrupts the target program to handle this exception immediately. 
\FEHSOSP{Moreover, the kernel also manages the context for each application thread to prevent the need for restarting the application from scratch after scheduling.}
Inspired by these, \sys{} introduces a comparable mechanism for robotic model inference.
\underline{\emph{First}}, we propose \sysctx{}, which stores the model execution state for each action (\textsection\ref{sub:design:ctx}). 
To mitigate storage overhead, we construct a context pool along with a dedicated eviction strategy. 
\underline{\emph{Second}}, we introduce the concept of \sysexcept{} for robotic actions (\textsection\ref{sub:design:except}). 
It can promptly interrupt the current action slice when any software- or hardware-defined exception is triggered, 
analogous to the kernel's exception handling. 
\underline{\emph{Third}}, to manage repetitive tasks, \sys{} employs the \sysreplay{} mechanism (\textsection\ref{sub:design:replay}), 
which allows the replay of robotic actions in a similar, albeit not identical, environment using the prior \sysctx{}.
The whole process is shown in~\autoref{fig:concepts:full-new}.
}

\subsection{\sysctxL{}}
\label{sub:design:ctx}
\begin{figure}
    \centering
    \setlength{\belowcaptionskip}{-20pt}
    \includegraphics[width=\linewidth]{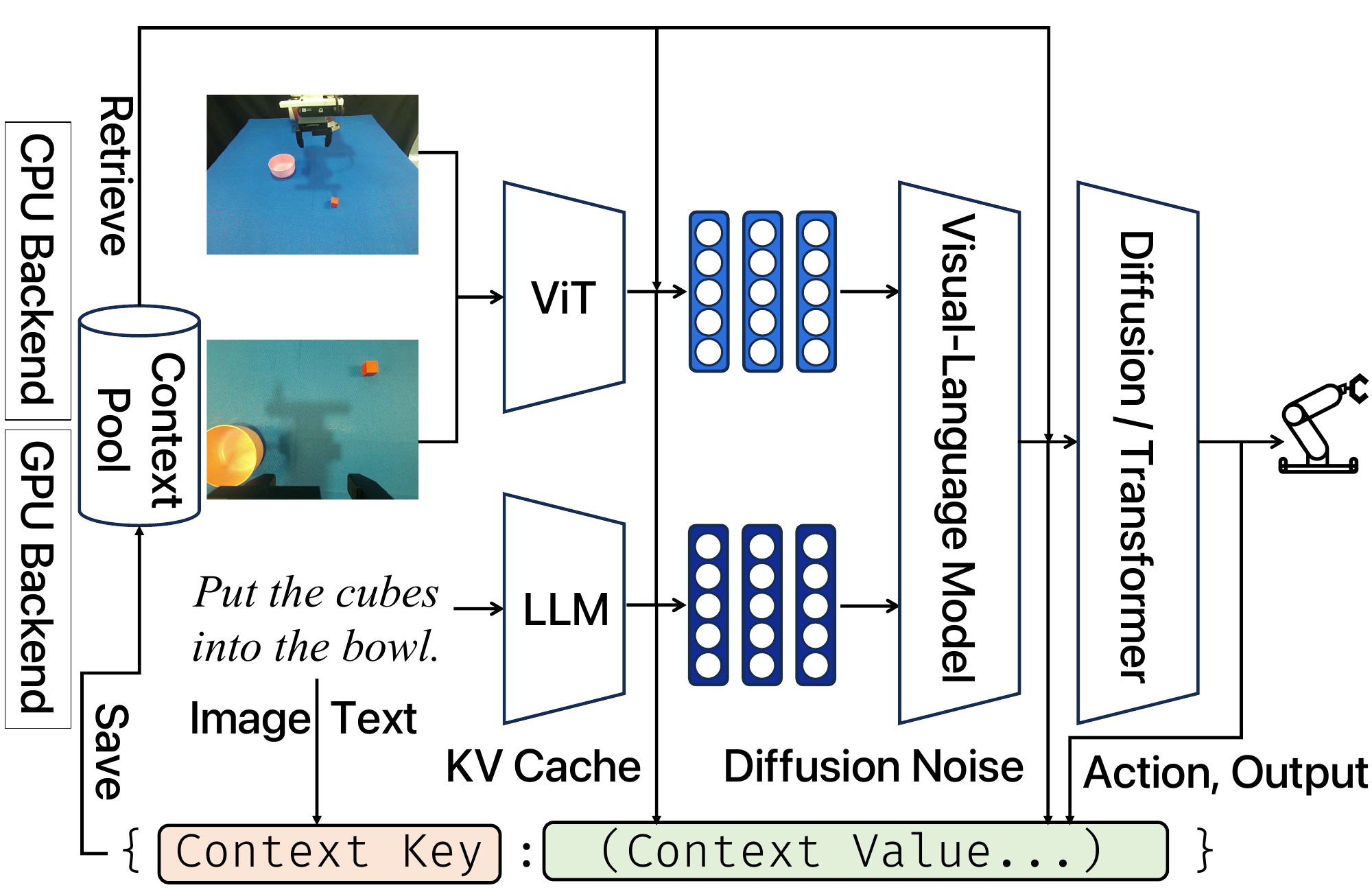}
    \caption{\textbf{\FEHSOSP{The structure of action context for multi-layer storage and retrieval.}}}
    \label{fig:concepts:context:full}
\end{figure}

The action context is used to represent a complete action for one task. 
It is a layered storage module that not only stores memories of action execution for repeatability but also speeds up subsequent reasoning processes.
The structure of action context is shown in~\autoref{fig:concepts:context:full}.

\subsubsection{Constructing the Action Context}
To efficiently reuse information from each inference, \sys{} stores all cacheable objects in the action context pool. 
This includes intermediate states during inference and the action trajectories produced by inference. 
Specifically, the intermediate states consist of the KV Cache from the Transformer model in visual recognition, latent vector generated at each time step by the Diffusion model for joint action generation, and the KV Cache from the Transformer model in the language understanding module. 
These cached intermediate states can be reused in future inference, thus reducing computational overhead and speeding up the time required for a single inference. 
The inference output includes data from the robot's joints and the output embedding obtained from the final inference. 
These action trajectories can guide the model to repeat previously executed actions in future tasks, enhancing planning efficiency and action consistency.

During each inference, the system first checks the action context pool to see if any similar and reusable intermediate states exist \REFINE{by comparing the similarity of the instruction and images}. 
If a reusable intermediate state \REFINE{like KV cache} is found, the system directly retrieves and uses these states to avoid redundant computations. 
If similar task trajectories are found, the system uses these previously inference results to guide the model \REFINE{by reusing latent vectors of successful action as the start state.}
This encourages it to mimic these trajectories in current inference. 
\REFINE{These approaches reuse previously cache and provide guidance for task completion, both increasing success rate and efficiency.}
\REFINE{After one single inference, \sys{} adds the unseen vector including new KV value and latent vector into the context pool for further use.}

\subsubsection{Layered Storage of Action Context}\label{lab:design:layered}
Saving all cacheable items would consume vast amounts of GPU memory, which is unacceptable for robotic applications that require real-time processing. 
However, the frequency of data usage are not uniform. 
As shown in~\autoref{tab:design:cache-size}, the KV cache from the environment recognition vision transformer and LLM KV cache occupies almost one-third of data but is accessed less frequently than the diffusion noise and output embeddings. 
Therefore, \sys{} manages data based on their access characteristics by prioritizing storage resources for caches that are frequently used but small in size, while applying optimized storage strategies for caches that are infrequently used but large in size. 

\begin{table}[!ht]
\centering
\caption[]{\textbf{Storage occupation for different cacheable items for one single action step.}}
\label{tab:design:cache-size}
\begin{tabular}{c|cc|c}
\toprule
Cacheable Items & $\pi_0$ & CogACT & Use Frequency \\ \midrule
ViT KV Cache & 63MB & 51MB & Low \\ \hline
LLM KV Cache & 165MB & 134MB & Medium \\ \hline
Diffusion Noise & 165MB & 301MB & High \\ \hline
Output Action & 14Byte & 14Byte & High \\ \bottomrule
\end{tabular}
\end{table}

Specifically, \sys{} adopts a hierarchical storage architecture to manage the action context. 
First, we determine the storage location of data based on its use frequency. 
\REFINE{For data that is frequently accessed, we store it in GPU memory to minimize data access latency and improve system performance. 
Conversely, for data that is infrequently accessed, we apply an on-demand recomputation strategy or move it to CPU.
Additionally, we do not place data that does not participate in computation near the computing components. }

\subsubsection{\sysactL{} Compression}
To minimize the storage space occupied by the action context, \sys{} employs a compression algorithm to reduce the size of the action context.
\REFINE{We observed that a single task often includes many repeated actions across different inference instances. 
For example, the paths for picking up objects might be very similar when the arm is exactly on top of the objects.}
Thus, we first build the action index for quickly retrieve the action and its corresponding vector and then introduce a ``virtual action'',  
which links repeated data with only one copy to reduce redundant storage. 
Then we perform auto eviction policy to reduce the size of the action context pool on GPU.

\myparagraph{Virtual Action.}
Different action sequences and environments often share common parts that can be reused, 
such as a next-step action of same task at the same location. 
Thus, \sys{} introduces a virtual action mechanism to enable cross-context action reuse, similar to the page table concept in operating systems.
To achieve this, \sys{} first creates an action index for each action to facilitate action indexing.
Once \sys{} collects an input, it serializes that input into a binary form in the background and calculates hash values for each segment of that binary data. 
In the first phase of hashing, \sys{} independently calculates a hash value for each segment using parallelism. 
In the second phase, it merges all the segments' hash values to generate the final hash result, serving as the unique index for action lookup and management. 
The process is listed in Algorithm~\autoref{lst:impl:index}.
To minimize the complexity, the hash functions involved are modular calculation.
Since this process is carried out asynchronously, it does not interference the normal operation of foreground tasks, thereby enabling fast action search and reuse without adding extra overhead.

\begin{algorithm}[!h]
\caption{Two-Phase Hashing for Action Index}
\label{lst:impl:index}
\begin{algorithmic}[1]
\REQUIRE 
\begin{itemize}
    \item Binary input data $B$ of length $N$.
    \item Segment size $s$ (number of bytes per segment).
    \item Hash functions $\mathit{HashFunc1}$ and $\mathit{HashFunc2}$.
\end{itemize}
\STATE \textbf{Phase 1: Segment-level Hashing}
\begin{enumerate}
    \item Partition $B$ into $k = \lceil \frac{N}{s} \rceil$ segments: $B_1, B_2, \dots, B_k$, \\
          where each $B_i$ has size $s$ bytes (except possibly the last one, if $N$ is not divisible by $s$).
    \item \textbf{for} $i = 1$ to $k$ \textbf{do}
    \begin{enumerate}
        \item $h_i \leftarrow \mathit{HashFunc1}(B_i)$
        \item Store $h_i$ in a list $H = [h_1,h_2,\dots,h_k]$
    \end{enumerate}
\end{enumerate}
\STATE \textbf{Phase 2: Aggregation}
\begin{enumerate}
    \item Combine all segment-level hashes from $H$ using $\mathit{HashFunc2}$:
    \[
        H_{\text{final}} \leftarrow \mathit{HashFunc2}(h_1, h_2, \dots, h_k)
    \]
    \item \textbf{return} $H_{\text{final}}$
\end{enumerate}
\end{algorithmic}
\end{algorithm}

Based on the index, \sys{} introduces ``virtual action'' mechanism that minimizes repeated storage 
by treating the context as a series of uniquely numbered actions with independent reference counts. 
These actions are stored in a contiguous area. 
Each context only keeps track of the actions it needs by referencing their IDs. 
When a new context adds an action, that action's reference count increases.
When a context is evicted or deleted, its referenced actions each have their reference count decreased. 
Any action whose reference count drops to zero is removed. 
This design eliminates unneeded storage overhead and boosts the \sys{}'s overall reuse efficiency.

\myparagraph{Auto Eviction.}
As stated in \S \ref{lab:design:layered}, different types of data occupy memory in different amount and exhibit different access frequency. 
\REFINE{To ensure transparent management of GPU memory and CPU memory resources,} 
\sys{} employs a hybrid cache management strategy that combines least recently used (LRU) with a priority-based approach, using dynamic cache eviction to improve resource utilization.
Specifically, \sys{} keeps a access counter for each vector, storing all vector on the GPU when sufficient space is available. 
If GPU memory becomes scarce, \sys{} automatically initiates auto-eviction, giving priority to objects with the smallest effect on outcomes or the lowest computational cost, and moving the least-used portion of data back to CPU memory. 
\REFINE{For example, the KV Cache in vision models can be easily recalculated. The extra cost is less than 10\%. When GPU resources are low, \sys{} will first evict the KV Cache of vision models.}
By balancing data distribution across different tiers and optimizing the eviction policy, \sys{} boosts memory efficiency and overall performance without sacrificing the response speed and accuracy for robots.

\subsection{\sysexceptL{}}
\label{sub:design:except}

\subsubsection{Exception}
During the execution of real robots, errors may occur.
Subsequent inference steps of the model can only correct errors produced by the algorithm in the model's output after the current slice has finished executing.
However, some errors that cannot be corrected or addressed in a timely manner through model inference.
We refer to these errors as \textbf{exceptions} and categorize them into two types: hardware exceptions and software exceptions.
These two types of exceptions are shown in~\autoref{tab:design:exception}.

\begin{table}[!ht]
    \caption{\textbf{Exception types defined in \sys{}.}}
    \label{tab:design:exception}
    \begin{tabular}{c|c|l}
    \toprule
    \# & Type & Exception \\ \midrule
    1 & \multirow{5}{*}{Hardware} & Collision. \\ \cline{1-1} \cline{3-3} 
    2 &  & Unreachable robot state. \\ \cline{1-1} \cline{3-3} 
    3 &  & Robot crash. \\ \cline{1-1} \cline{3-3} 
    4 &  & Too big torque. \\ \cline{1-1} \cline{3-3} 
    5 &  & Too big angular momentum. \\ \midrule
    6 & \multirow{2}{*}{Software} & Not expected action. \\ \cline{1-1} \cline{3-3} 
    7 &  & Software defined condition violation. \\ \bottomrule
    \end{tabular}
\end{table}

\myparagraph{Hardware Exception.} 
The hardware exception is defined as the robot's abnormal mechanical state.
\REFINE{The model is unaware of these abnormal mechanical state because it did not take such abnormal state as input to model inference. }
To address this, \sys{} monitors the robot's state at runtime,
converting the robot's corresponding abnormal state into a hardware exception and 
raising it to the model to immediately interrupt the execution.
Specifically, when \sys{} detects that the physical hardware returns an abnormal state,
it immediately interrupts the remaining actions in the \sysslice{} that have not been executed,
stopping the execution of the current \sysslice{}.

\myparagraph{Software Exception.}
\REFINE{Software Exception means the unexpected state in the robot's execution defined by software.}
It has larger scope than hardware exception.
For example, when controlling the gripper for grasping, the gripper may occasionally close without actually gripping anything, 
resulting in a failed grasp. 
\REFINE{Although in a complete \sysslice, a single failed action may not affect the normal execution of subsequent actions, 
continuing to operate after an uncorrected software exception can further cause hardware exceptions that are difficult to handle.}

\begin{table}[!h]
    \caption{\textbf{Example check of software defined condition.}}
    \label{tab:design:fault-condition}
    \begin{tabular}{c|c}
    \toprule
     Action & Expected Outcomes \\ \midrule
     Gripper grasping & Gripper’s minimum gap exceeds 0.\\ \hline
     Move stick & Change in the stick’s angle. \\ \hline
     Move manipulator & Change in the manipulator’s angle. \\ \hline
     Apply force & Force sensor exceed 0. \\ \hline
     Send stop & All actuators halt. \\ \bottomrule
    \end{tabular}
\end{table}

Following the principle that each action should produce an observable effect on the environment, 
\sys{} defines a set of expected outcomes for each action.
For example, when the gripper is grasping,
the expected outcome is that the gripper's minimum gap exceeds 0.
\REFINE{Such expected outcomes can be rule-based or defined from the LLMs.
Examples of the expected outcomes is shown in~\autoref{tab:design:fault-condition}.}

\begin{figure}[!ht]
    \centering
    \includegraphics[width=\linewidth]{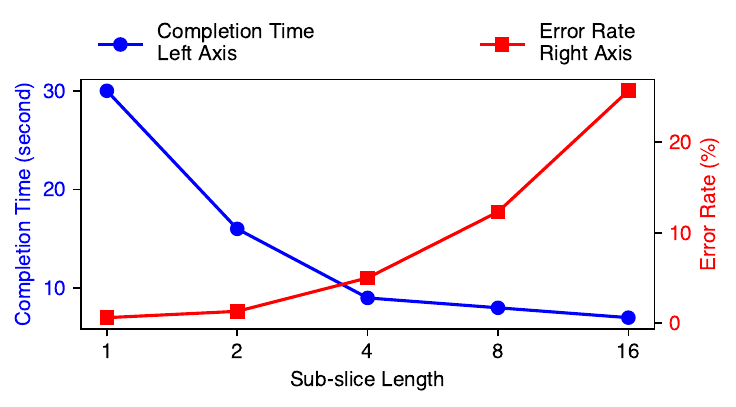}
    \caption{\textbf{Trade-off between checking frequency and accuracy of the action.}}
    \label{fig:design:software-exception:tradeoff}
\end{figure}

However, raising software exception is not as straightforward as hardware exception 
since it requires additional time on the critical path of action to determine whether the action is successful or not.
\FEHSOSP{If the system performs checks more frequently, the proportion of correct actions will increase; 
however, this will also lead to slower operation of the robot.}
There is a trade-off between the checking frequency and accuracy of the action, as shown in~\autoref{fig:design:software-exception:tradeoff}.

To address this, \sys{} introduces a mechanism that divides the action into smaller sub-slices.
\REFINE{After each sub-slice is executed, \sys{} asynchronously checks the environment gathered at the end of sub-slice to see whether the relevant robot joint states match the expected states. }
All output checks only check whether the action has taken effect, 
rather than determining whether the action is semantically correct or meeting its intended goal. 
The length of one sub-slice is not always fixed.
\REFINE{Its length is related to the current action intent.
If the robot shows a very large range of movement,}
\sys{} will reduce the length of the sub-slice.
The final length of the sub-slice is determined to be between 2 to 5 steps.
If a sub-slice is judged to have failed, 
\sys{} will immediately raise a software \sysexcept{} to the model.


\subsubsection{Handler}
The exception handler of \sys{} is used to recover from the abnormal state of the robotic arm.
Due to the different types of exceptions, \sys{} will take different exception handlers.
\REFINE{Among these exception handlers, there are common parts and specific parts.}

\myparagraph{Common Part.}
The common part is that \sys{} immediately interrupts the execution of the current action slice.
\sys{} immediately collects and updates the current environment data and sends it to the GPU to start a new inference. 
Meanwhile, to prevent previously erroneous actions from polluting future operations, 
\sys{} discards any sub-slices that have not yet been executed. 
It then leverages a newly generated inference result,
using the updated environment data to generate subsequent actions. 
This approach allows the model to promptly correct errors and maintain stability and consistency throughout the execution process.

\myparagraph{Specific Part.}
The specific part is that \sys{} will handle the exception differently based on the type of exception.
For software exceptions, \sys{} just clear the current action slice and do not reset the robot power state.
For hardware exceptions, \sys{} will reset the robot to a safe state first.
Then, \sys{} will decide whether to restart the robot based on the severity of the problem and exception number.
For example, if the current action causes too big torque (Exception \#4) or too big angular momentum (Exception \#5), \sys{} will not restart the robot.
If the robot's mechanical arm collides (Exception \#1), \sys{} will restart the robot.

\begin{figure}
    \centering
    \includegraphics[width=\linewidth]{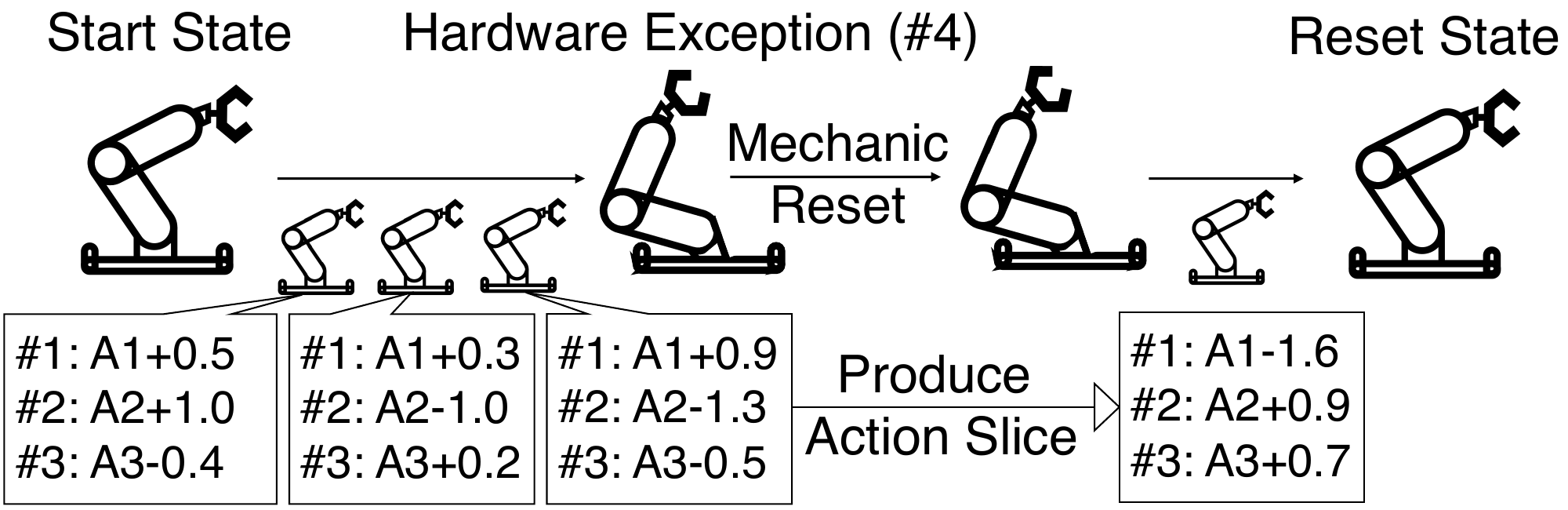}
    \caption{\textbf{Robot physical state reset. }A1, A2, A3 represent the different ankle of the model. When synthesizing action slices, computation is based on robot configuration and might be different from direct addition. \#1, \#2, \#3 represent the different action atomization operations.}
    \label{fig:design:state-reset}
\end{figure}

\FEHSOSP{However, resetting the robot to a safe state cannot be achieved by merely recording every movement trajectory of the robotic arm and rolling back these trajectories, 
as the movement paths often contain a significant amount of redundancy.}
\sys{} designs a smaller atomic action unit to represent the robot's physical state changes called \textbf{action atomization}.
Different from the traditional action chunking approach,
action atomization limit the action to only 1 degree of freedom (DoF) of the robotic arm and it can deconstruct continuous movements into smaller atomic action units that represent micro-angular displacements in joint space and displacement vectors in the coordinate system.
\REFINE{During rolling back, \sys{} follows he concept of spatial composite vectors to combine these atomic actions into a shorter action slice.}
To implement this mechanism, \sys{} maintains a buffer on the CPU to record these actions. 
When the buffer is full, \sys{} automatically consolidates all actions into one action slice. 
The entire process is shown in~\autoref{fig:design:state-reset}.

\subsection{\sysreplayL{}}
\label{sub:design:replay}
\REFINE{Considering a long-horizon task with multiple repetitive jobs, 
when the model continuously outputs meaningless actions, 
\sys{} will trigger an replay signal to prevent the inference from hanging and perform a replay to guide the model to infer based on previously successful experiences.}
Different from traditional operating systems,
when meaningless actions are detected, \sys{} cannot directly reuse the previous \FEHSOSP{trajectories} to resume execution.
This is because the current environment and the environment in the context are not exactly the same,
which means the steps produced for the last context may not be valid anymore.
To address this, \sys{} adopts a hybrid approach that combines inference with replay, combining state reset and action regeneration.

\subsubsection{Replay Signal}
\REFINE{\sys{} introduces replay signal to indicate the robot is actually hanging and requires additional control. 
In the traditional operating system, SIGNAL is triggered to force the kernel to reclaim CPU control and switch to other processes/tasks.
This prevents a single process from hanging the CPU and exhausting all resources.}
In robotic models, sometimes the model do not generate any moving action for a long time, making the arm keep idle while the task is not completed.
This is similar to the CPU being occupied by a single process for a long time.
\REFINE{Specifically, if all actions within an action slice are marked as completed, or if no actions are executed while the task remains unfinished, 
\sys{} will send a replay signal.}
This prevents the inference process from hanging indefinitely.
\FEHSOSP{Subsequently, \sys{} incorporates a description of the unfinished task into the prompt 
and utilizes the vector from the action context in the context pool as the initial vector for the immediate new inference, 
rather than employing a random vector.}

\subsubsection{Action Regeneration}
\REFINE{\sys{} uses replay to stop meaningless actions and use previous successful experiences to guide further inference.
However, previous experience cannot be directly applied due to the changing of the environment, as well as factors such as the position and orientation of target objects. }
When the environment undergoes slight changes or target object positions shift, action trajectories that were previously successful are no longer applicable. 

To effectively address this issue and enhance the robustness of action replay, 
\sys{} employs a guided inference strategy for the action generation using a diffusion-based action generation model, 
instead of directly reusing previously successful contextual action data as direct output. 
Specifically, \sys{} uses successful action vectors validated through prior environmental interaction as the initial vector for the current action diffusion process, 
which is equal to performing an additional inference of diffusion steps based on the existing context. 
This method explicitly guides the model during the initial phase of action generation, making the generated trajectory more likely to leverage previously effective action features, 
thereby increasing success rates in similar environments that have undergone slight modifications.
Additionally, to ensure the adjusted action sequence adapts to changes in the environment and target objects, 
\sys{} will also changes the prompt to include the current environment and task description.
The added prompt will clearly direct the model to reanalyze the current environment and formulate new action slice,
rather than simply reusing previous conclusions or encoded information. 
With this enhanced prompting mechanism, the LLM is able to proactively utilize the currently perceived state of the environment to re-engage in inference and action generation.

%% file: src/4-impl.tex
\section{\sys{} Implementation}
\begin{figure}[!h]
    \centering
    \includegraphics[width=\linewidth]{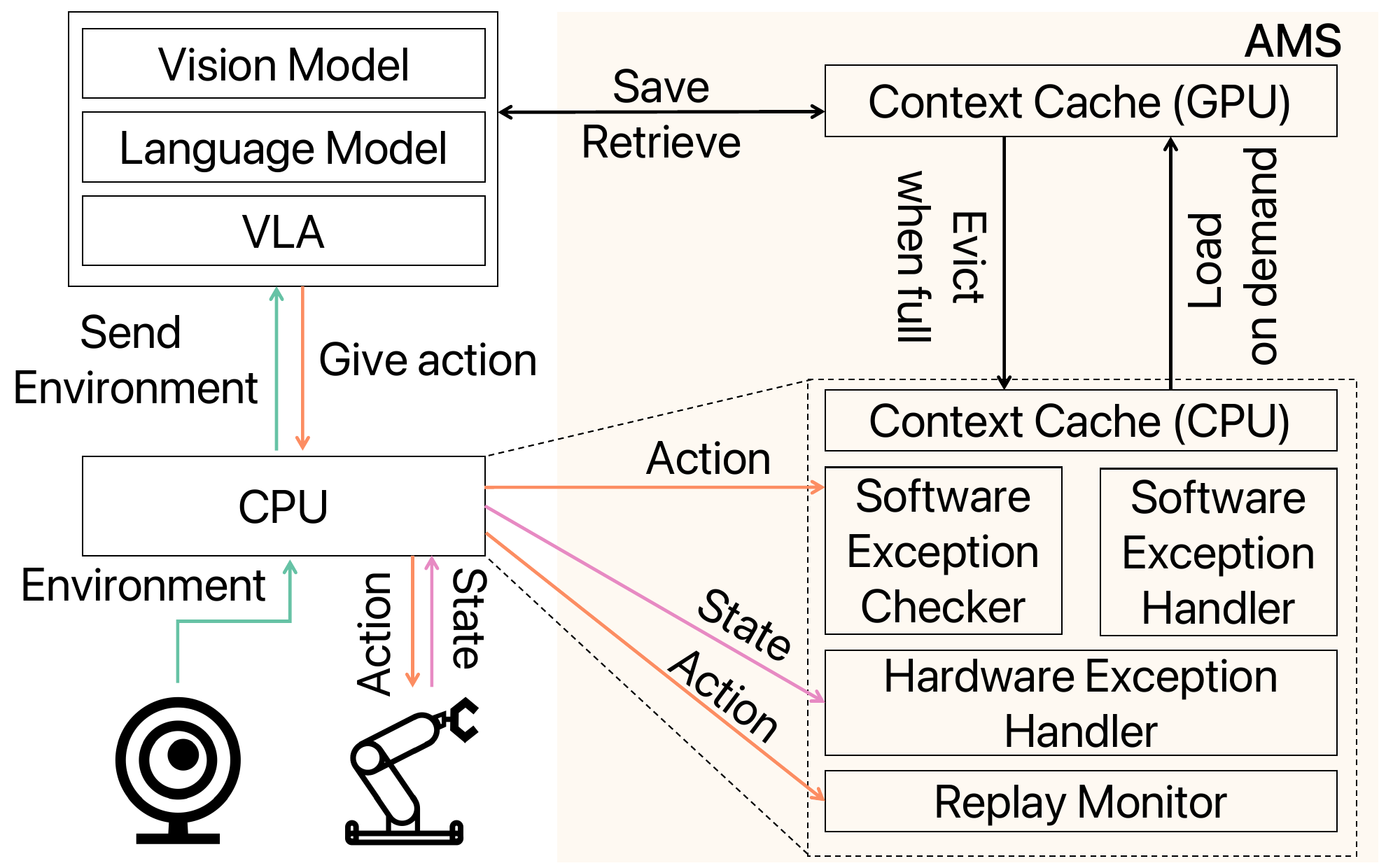}
    \caption{\FEHSOSP{\textbf{Implementation of \sys{}.} The \sys{} system contains several key components: 
    Action context pool manager, \sysexcept{} handler and \sysreplay{} signal trigger (replay monitor in the figure).}}
    \label{fig:impl:overview}
\end{figure}

\sys{} is a module within the framework.
It contains a static auto hooking tool and a runtime library. 
To maintain maximum compatibility with existing systems and to minimize interference with the inference process, \sys{} scans the code and inserts hook functions to locate the action slice by static analysis. 
For the runtime library, \sys{} include the context cache, the exception handler, and the replay signal trigger.
We place the whole control logic of \sys{} on the CPU.
For context manager, we prioritize putting action context on the GPU. 
If the GPU context cache runs out of space, \sys{} evicts vectors which are less frequently used and have minimal impact on inference results. 
For exception handler, \sys{} will monitor if the action responded as expected to the action sent to the robot.
If not, \sys{} will immediately stop sending further instructions, reset the robot and re-collects environment data to initiate a new inference. 
For action replay signal trigger, when the model finally outputs a stop signal, \sys{} will first check whether a replay should be applied.
If \sys{} requires replay, the replay signal is up. Then, \sys{} will block the stop signal and retrieve stored contexts from the context pool one by one to guide subsequent inferences, helping avoid workflow disruptions caused by misjudgment or intermediate errors.
The whole process is shown in~\autoref{fig:impl:overview}.

%% file: src/6-eval.tex
\section{Evaluation}

To show the performance of \sys{}, we conducted both real-world robotics and simulation with the models below. 
We seek to answer the following three questions.

\begin{itemize}
    \item Can \sys{} accelerate the \FEH{task} completion? (\S \ref{c5:efficiency})
    \item Can \sys{} improve the \FEH{task} success rate? (\S \ref{c5:accuracy})
    \item How does the \sys{} perform in comparison to traditional algorithm-based methods? (\S \ref{c5:algorithm})
\end{itemize}

\myparagraph{Setup.}
We selected real-world robotics (JAKA s5~\cite{Home68:online}) and simulation robotics (WidowX~\cite{HomeTros59:online} in SimplerEnv~\cite{li24simpler}) as our testbed.
We use three SOTA models, Octo~\cite{team2024octo}, CogACT~\cite{li2024cogact} and $\pi_0$~\cite{black2024pi0visionlanguageactionflowmodel}, to evaluate the performance of \sys{}.
\FEH{To evaluate the generalization capabilities of the models, 
these models are not fine-tuned for our test jobs; instead, they are pretrained to adapt to our robotic platforms.}
We use direct inference without \sys{} and VLA-Cache~\cite{xu2025vla} as our baseline.
\REFINE{We use three representative tasks from Basic Manipulations and Multiple Object Interactions of RoboMind dataset~\cite{wu2024robomind}, which is consistent with algorithmic solution evaluation~\cite{wang2024rise}.}
\FEHSOSP{Given that there may be multiple objects on the desk, these long-horizon tasks can be structured to several repetitive jobs for different objects.}




\myparagraph{Metrics.}
To evaluate the performance and efficiency of \sys{}, we report the success rate, average finish step and end-to-end latency.
Success rate refers to the ratio of the number of times a robot successfully completes a task to the total number of trials. 
Average completion step refers to the average number of action steps required for the robot to successfully complete a task, excluding the first execution. 
End-to-end latency refers to the time required for one successful execution.
It depends on the time taken for each step and the number of steps required.
We set a limit of 1500 steps for each task, which is twice the size needed to complete a single task.

\begin{figure}[!h]
    \centering
    \includegraphics[width=\linewidth]{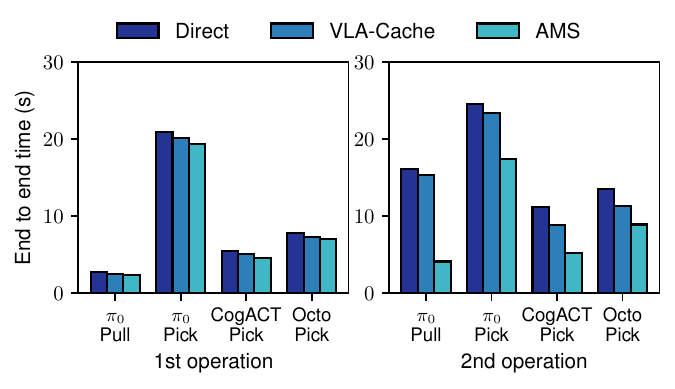}
    \caption {\textbf{End-to-end efficiency evaluation for \FEH{the first and second time execution}. }}
    \label{fig:eval:end2end:time}
\end{figure}

\subsection{End-to-end efficiency evaluation}\label{c5:efficiency}
We tested the \sys{} for end-to-end execution time in both real-world and simulation scenario. 
The overall end-to-end efficiency evaluation is shown in~\autoref{fig:eval:end2end:time}.

\FEH{The total execution time is related to the Actions Per Second (APS) metric and the overall number of action steps required to complete the task. 
The overall APS is defined as the minimum value between the maximum hardware APS of the robot and the APS of the model inference. 
In practical scenarios, the hardware APS of the robot is usually lower than that of the model in most situations. 
Therefore, enhancing the model's APS exerts a lesser influence on the results compared to reducing the number of action steps.
To comprehensively evaluate \sys{}, we measured the execution time for both the first and second executions of the robotic task. 
In the first execution, \sys{} achieves a modest acceleration, reducing the execution time by 7\% to 15\%. 
This limited improvement is attributed to the fact that the cache can enhance model inference but does not decrease the total number of action steps during the first execution.}
For the second execution, \sys{} shows significant acceleration due to the reused context, reducing execution time by 29\% to 74.4\%. 
This is because the combination of \sysctx{} and \sysintr{}, which allows for direct use of cached previous context information during inference and reduces invalid hanging steps.

To show why \sys{} can accelerate the execution of robotics, 
we breakdown the performance with the effects of the number of total steps and the actions per second of \sys{}.

\begin{figure}[!h]
    \centering
    \includegraphics[width=\linewidth]{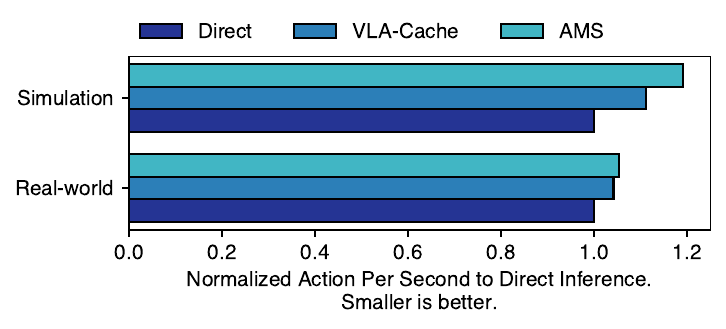}
    \caption{\textbf{Actions Per Second test. }Evaluate the overall APS of the robotic system with and w/o \sys{} support.}
    \label{fig:eval:end2end:perstep}

\end{figure}

\myparagraph{APS (Actions Per Second)}
We measured the APS to test how \sys{} accelerates individual step inference. 
The results are shown in~\autoref{fig:eval:end2end:perstep}. 
\FEH{In real-world environments, the inherent limitations of hardware in robotic arms restrict the overall performance. 
Accelerating model inference speed has a minimal impact on the actual speed of robotic actions; 
therefore, \sys{} only achieves an average improvement of 5\%.}
\FEH{In simulation environments or future robots equipped with higher hardware APS, 
the acceleration effect achieved by increasing the model inference APS becomes increasingly significant. 
\sys{} demonstrates a 19\% improvement in single-step performance compared to the baseline model, 
as it effectively manages the \sysctx, which reduces redundant computations for the KV cache and latent vectors in VLA-based robotic models.}

\begin{figure}[!h]
    \centering
    \includegraphics[width=\linewidth]{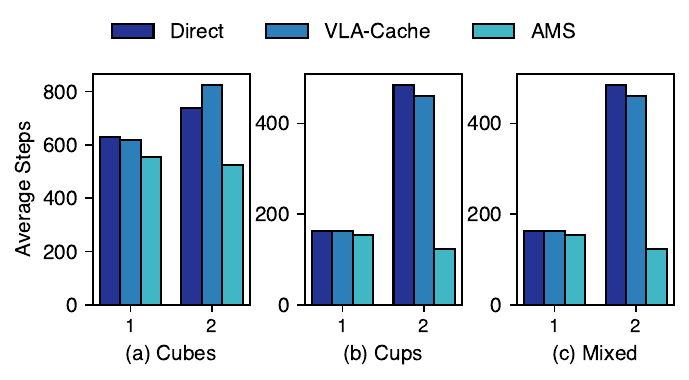}
    \caption{\textbf{Execution steps of $\pi_0$ model for different tasks in real-world scenario.} ``Cubes'' represents picking all cubes and placing into the bowl. ``Cups'' represents pulling down all cups. ``Mixed'' represents arranging the whole desk which contains cubes and cups.}
    \label{fig:eval:end2end:real:steps}
\end{figure}

\begin{figure}[!h]
    \centering
    \includegraphics[width=\linewidth]{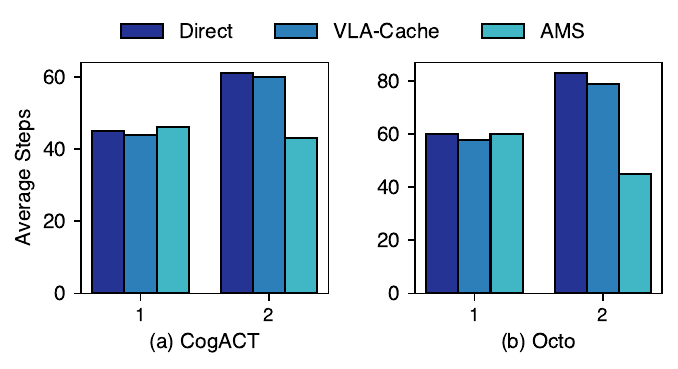}
    \caption{\textbf{Execution steps in the simulation scenario with CogACT and Octo model for picking the spoons.}}
    \label{fig:eval:end2end:simulation}
\end{figure}

\myparagraph{The number of steps}
\FEH{Another factor that influences end-to-end efficiency is the total number of action steps.}
We conducted steps count for both real-world and simulation case, and the result is shown in~\autoref{fig:eval:end2end:real:steps} and ~\ref{fig:eval:end2end:simulation}.
For the real-world case, \sys{} sustains excellent performance compared to direct inference.
In the number of steps needed, the \sys{} system reduces the steps by 29\% for picking the cubes and 74.4\% for pulling the cups in the second step compared to a inference without \sys{}. 
Additionally, compared to the first execution of the task using \sys{}, 
the average number of steps in the second execution decreased by 5.7\% for picking up the cubes and by 20\% for pulling the cups.
For the simulation case, \sys{} also performs significant steps improvement.
Compared to the baseline model using direct inference, \sys{} can reduce the number of steps required to complete the task by 29\% to 45\%, and compared to the solution using VLA-cache, \sys{} can reduce the number of steps by 28\% to 43\%.
\FEH{This performance improvement is attributed to the caching and reuse of the latent vector for VLA models, 
which enables the robot's subsequent actions to align with previous trajectories, thereby minimizing hanging and disorderly actions.}

\subsection{Task success rate evaluation}\label{c5:accuracy}
Another important factor affecting the model's performance is the task success rate.
\FEH{To evaluate the generalization capability of repetitive/similar tasks for \sys{}, 
we test the robot on tasks that are both within and outside the training dataset.
In this section, the original model is trained on the manipulation of a single cup or cube, 
yet it is expected to handle multiple objects during operation.}

\begin{figure}[!ht]
    \centering
    \includegraphics[width=\linewidth]{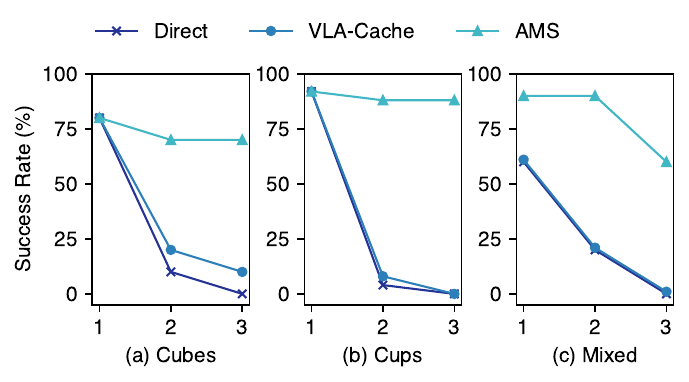}
    \caption{\textbf{Success rates of $\pi_0$ model for different tasks in real-world scenario.} ``Cubes'' represents picking all cubes and placing into the bowl. ``Cups'' represents pulling down all cups. ``Mixed'' represents arranging the whole desk which contains cubes and cups.}
    \label{fig:eval:success-rate:real}
\end{figure}

In real-world scenarios, the \sys{} can achieve 7$\times$ in success rate for desk arranging task via picking.
It can also achieve up to 24$\times$ increase in accuracy for single-arm movement job via pulling. 
As shown in~\autoref{fig:eval:success-rate:real}, for objects more than two or for different shaped objects, \sys{} can still improve the task success rate significantly.

\begin{figure}[!ht]
    \centering
    \includegraphics[width=\linewidth]{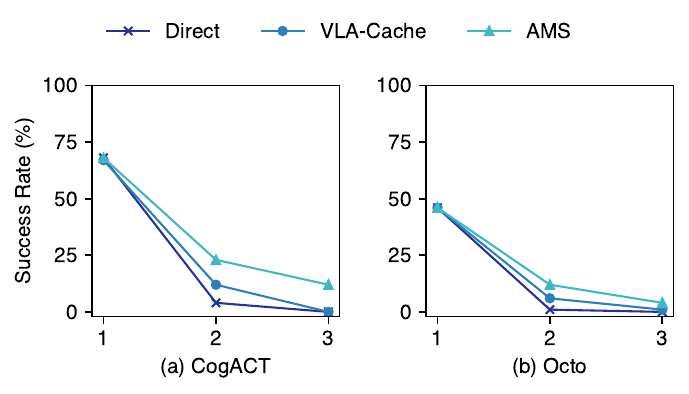}
    \caption{\textbf{Success rates in simulation scenario with CogACT and Octo model for picking the spoons.}}
    \label{fig:eval:success-rate:sim}
\end{figure}

In simulation environment, the success rate of the task is not as high as in the real-world environment, 
but compared to the baseline model using direct inference, \sys{} still achieves a 5 to 12$\times$ improvement in accuracy, 
and a 2$\times$ performance improvement compared to the baseline model.
The results are shown in~\autoref{fig:eval:success-rate:sim}.
\FEH{We analyzed the execution process of the task and identified that this drop in success rate is primarily attributable to two factors. 
First, the simulation environment is unable to detect all hardware exceptions. 
Consequently, \sys{} cannot promptly trigger an interrupt for fault actions, which may cause the error propagation. 
Second, the limited space within the simulation environment increases the interference from other objects during model inference, 
in contrast to the real-world environment.}



\subsection{Case Study: Picking two cubes on the desk}\label{c5:case}

\begin{figure*}[!tbp]
    \centering
    \includegraphics[width=\linewidth]{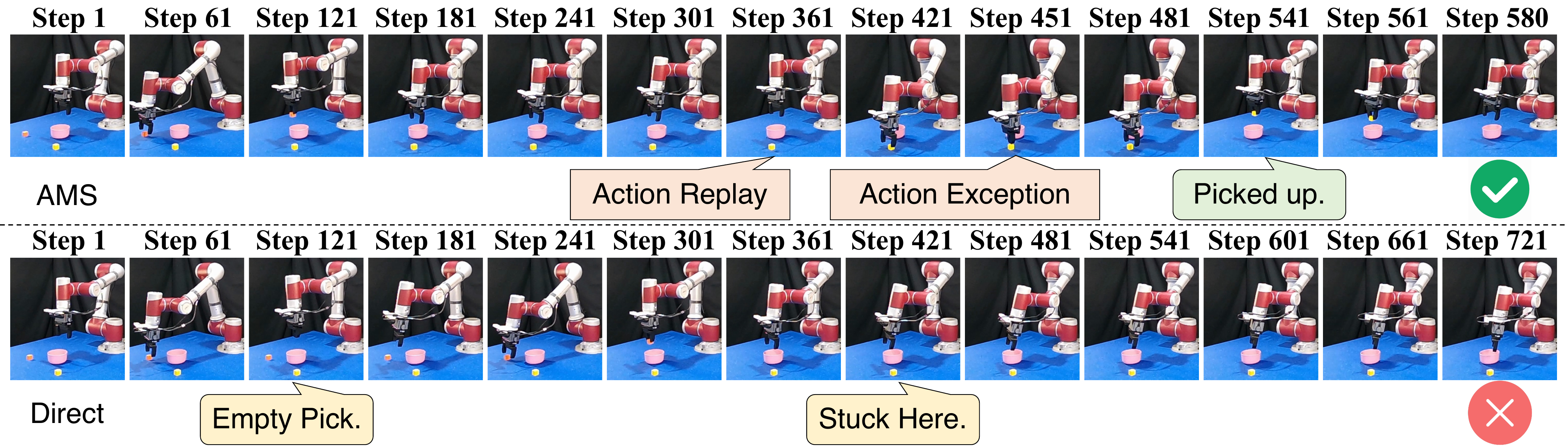}
    \caption{\textbf{Model output action for ``picking all cubes and placing into the bowl'' job by model $\pi_0$.}}
    \label{fig:eval:end2end:real:full:cube:example}
\end{figure*}

We further conducted a detailed step-level evaluation of \sys{}'s performance in object-picking tasks, as illustrated in~\autoref{fig:eval:end2end:real:full:cube:example}.
Completing the first task, which is picking up the first small cube, the baseline model and \sys{} performed similarly, both successfully accomplishing the task between steps 1 and 301. 
At step 361, \sys{} detected that all actions in the current action slice were producing no valuable effects, 
and \sys{} immediately triggered an replay signal and performed a context switch. 
This process was immediately reflected at step 361 as a system reset: the robotic arm returned to its initial position. 
\FEH{Subsequently, in the next action slice, \sys{} successfully reoriented towards the yellow cube using the prior \sysctx, allowing the task to continue.
By reusing the \sysctx, \sys{} can pick up the second cube more efficiently, requiring fewer action steps compared to the first pickup.}
In contrast, the baseline model became stuck, with no changes in its actions from step 361, unable to further progress the task, ultimately leading to failure.
\FEH{Furthermore, when \sys{} utilized the previous context for inference, it identified a fault at step 451. 
\sys{} promptly interrupted the subsequent actions and executed an exception handler to revert to step 446, 
ultimately enabling the successful completion of the entire task.}

\subsection{Comparison with algorithmic solutions}\label{c5:algorithm}
\FEH{We further compared \sys{} with algorithmic solutions, 
such as retraining the model to address a greater number of repeated tasks. 
Specifically, we evaluated the success rates of \sys{} against those achieved by using retrained models to accomplish varying numbers of tasks.
For single-arm movement job, \sys{} is compared with scenarios where the model was trained to pull down 1, 3 or 6 objects.
}


\begin{figure}[!h]
    \centering
    \includegraphics[width=\linewidth]{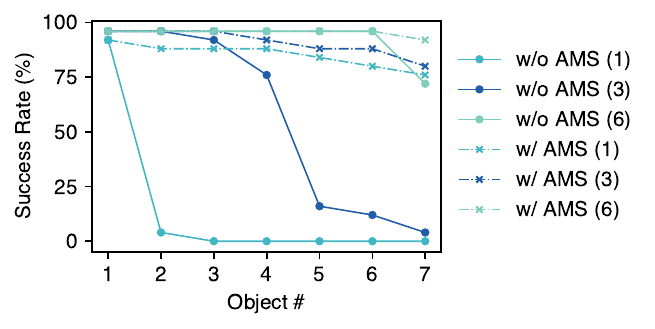}
    \caption{\textbf{Comparison between \sys{} and algorithm-based solution in real-world robots.} Number in the brackets  means the maximum repetition counts of pulling in the training set.}
    \label{fig:eval:generalization:cubes}
\end{figure}

The results in~\autoref{fig:eval:generalization:cubes} show that \sys{} effectively improves the model's ability to generalize when performing repetitive tasks. 
When the required number of repetitions exceeds the maximum included in the training data, 
the success rate of the model declines sharply. 
In the task, the model's accuracy decreases from 92\% to 16\% for \FEH{the original model or fine-tuned model (3 objects)}.
The model's accuracy also decreases from 96\% to 72\% for the \FEH{fine-tuned model with 6 objects}.
However, with \sys{} support, the model's accuracy only decreases by up to 8\%, 
indicating that \sys{} effectively maintains the model's accuracy as the number of repetitions increases.
\FEH{Furthermore, even when employing the original model (1 object), 
the \sys{} demonstrates a higher success rate compared to the fine-tuned model (6 objects) when attempting to pull the seventh cube (out of dataset),
and a comparable success rate when pulling the first to sixth cubes (in dataset).
}

\subsection{Ablation Study}\label{c5:ablation}
To demonstrate the impact of each module on the model's final inference results, we conducted an ablation study to show how different modules affect accuracy and the number of steps completed.

\subsubsection{Performance breakdown}
\begin{figure}[!h]
    \centering
    \includegraphics[width=\linewidth]{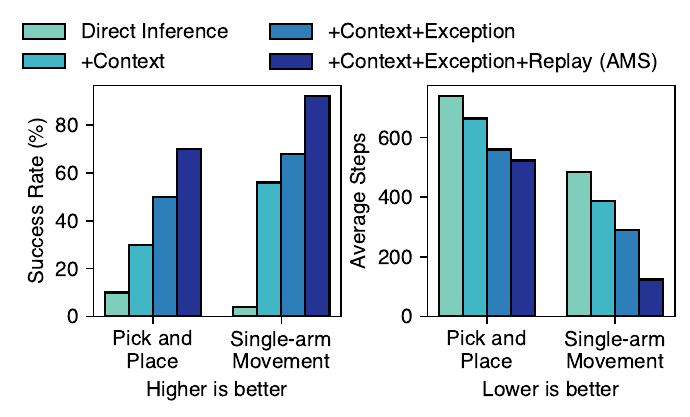}
    \caption{\textbf{Impact on success rate and average steps of \sys{}.} ``Single-arm movement'' is pulling down the cups.}
    \label{fig:eval:ablation:full}
\end{figure}

In this part, we will analyze impact of different parts on \sys{}. The results are shown in~\autoref{fig:eval:ablation:full}.

For correctness, \sys{}'s context reuse effectiveness varies by action type. In ``pick and place'' tasks, it improves success by 20\%, while in ``single-arm movement'' tasks, it boosts success by 48\%, quadrupling the original rate. 
After adding an action fault mechanism, ``pick and place'' success rose by 40\%, reaching 5 times the original rate, but ``single-arm movement'' only increased to 72\% due to already low error rates for such job.
With an action interrupt mechanism, \sys{} detects task termination signals and switches context, raising success rates to 70\% for ``pick and place'' and 92\% for ``single-arm movement'', increasing to 7 to 23 $\times$, respectively.

For efficiency, the main improvement of \sys{} in real-world environments comes from reducing the number of steps. 
The reduction varies by task. For pick and place tasks (like picking up cubes or arranging mixed objects), \sysctx{} can reduce steps by 10.1\%. For single-arm movement tasks (like pulling down cups), \sys{} can improve by 20\%. This is because \sysctx{} can more easily reuse previous experiences for pull tasks. Adding \sysexcept{} can further reduce steps by 24.2\% and 40.3\%, respectively, while adding action replay can reduce steps by 29.2\% and 74\%.




\subsubsection{Sensitivity of Replay Signal Threshold}
Judging when the model should replay is an important part for action replay.
The model does not output whether the current action is completed. We need to determine 
whether the current model has completed current task based on the current state of the robotic arm.
Therefore, we analyzed various parameters for raising replay signal and the results are shown in~\autoref{tab:eval:sensitivity:interrupt}.
We use a task that involves picking up and placing balls using a gripper. There are two balls on the table in total that need to be placed.

\begin{table*}[!ht]
    \caption{\textbf{Sensitivity of action replay signal threshold.} ``Arm'' refers to the rotation angles of each joint in the robotic arm, which are passed as parameters to control the robotic arm. ``Gripper'' refers to the minimum distance between the gripper fingers.}
    \label{tab:eval:sensitivity:interrupt}
    \centering
    \begin{tabular}{c|cc|c|cc|ccc}
        \toprule
        \multirow{2}{*}{No.} & \multicolumn{2}{c|}{Signal Threshold} & \multirow{2}{*}{\begin{tabular}[c]{@{}c@{}}Trial\\ Success\end{tabular}} $\uparrow$ & \multicolumn{2}{c|}{Average Steps} & \multicolumn{3}{c}{Replay Signal} \\ \cline{2-3} \cline{5-9} 
         & Arm & Gripper &  & 1st$\downarrow$ & 2nd$\downarrow$ & Total & False Positive$\downarrow$ & False Negative$\downarrow$ \\ \midrule
        1 & 1e-5 & 1 & 7 / 10 & \begin{tabular}[c]{@{}c@{}}555.5 (288 - 1387)\end{tabular} & \begin{tabular}[c]{@{}c@{}}523.7 (299 - 1010)\end{tabular} & 8 & 0 & 2 \\ \hline
        2 & 1e-4 & 1 & 6 / 10 & \begin{tabular}[c]{@{}c@{}}709.57 (304 - 1590)\end{tabular} & \begin{tabular}[c]{@{}c@{}}614 (576 - 602)\end{tabular} & 37 & 31 & 1 \\ \hline
        3 & 1e-1 & 1 & 6 / 10 & \begin{tabular}[c]{@{}c@{}}482.83 (317 - 654)\end{tabular} & \begin{tabular}[c]{@{}c@{}}449.67 (311 - 601)\end{tabular} & 40 & 34 & 0 \\ \midrule
        4 & 1e-5 & 0.5 & 6 / 10 & \begin{tabular}[c]{@{}c@{}}1129.25 (299 - 1659)\end{tabular} & \begin{tabular}[c]{@{}c@{}}515.8 (353 - 752)\end{tabular} & 4 & 3 & 7 \\ \hline
        5 & 1e-5 & 5 & 3 / 10 & \begin{tabular}[c]{@{}c@{}}691.28 (277 - 1470)\end{tabular} & \begin{tabular}[c]{@{}c@{}}322.3 (293 - 363)\end{tabular} & 49 & 43 & 0 \\ \hline
        6 & 1e-5 & 10 & 3 / 10 & \begin{tabular}[c]{@{}c@{}}1283.7 (250 - 3358)\end{tabular} & \begin{tabular}[c]{@{}c@{}}519.6 (354 - 652)\end{tabular} & 68 & 61 & 0 \\ \bottomrule
        \end{tabular}
        
\end{table*}

We tested that the setting of the replay signal threshold affects the model's success rate and average number of steps, and that the replay signal settings for different physical components are not the same.
We divided the test into two categories based on the type of action: joint movement (Case 1-3) and gripper movement (Case 4-6).
The setting of the threshold is not simple, and there is a trade-off between false positive and false negative.
When the model's threshold is set too low, it results in a low occurrence of false positives, but there may be some undetected false negatives.
When the model's threshold is set too high, it can almost avoid false negatives, but the occurrence of false positives is high, 
which can affect performance and even lead to an up to 50\% decrease in accuracy due to excessive replay action.



%% file: src/7-related-conclu.tex
\section{Discussion}
\myparagraph{Integrating with LLMs.}
Currently, \sys{} uses a numerical threshold method to determine replay signals.
Although this approach is straightforward and effective, it still demands significant human effort for setup and calibration. 
In the future, we plan to integrate \sys{} with LLMs to automatically identify and add new \sysexcept{} event classifications. 
Such integration requires no change to \sys{}.
Specifically, \sys{} will utilize LLMs and MLLMs during action to intelligently fuse multi-sensor data and semantic information, allowing for real-time judgment and decision-making on replay signals.

\myparagraph{Context Transfer capability.}
Relying on context alone doesn't help with unfamiliar tasks due to the crucial role of LLM and ViTs mapping in fine-tuning. 
Without a unified representation for new visual inputs and language instructions, the model struggles with unseen tasks. 
While \sys{} successfully enhances overall capabilities in long-horizon known tasks, improving performance on previously unseen scenarios demands targeted algorithmic modifications.

\myparagraph{Resource Limitation.}
\sys{} doesn't significantly raise computational costs during inference 
because it only caches intermediate states, needing no more memory than a single inference. 
More GPU memory allows us to store additional action sequences, enhancing efficiency by quick action reuse. 
Limited GPU memory might prompt extra CPU-GPU data transfers over PCIe, but these have minimal impact. 
As shown in~\autoref{fig:motiv:vla}, a single inference takes roughly 200 milliseconds, whereas executing these action takes about 1 second. 
Therefore, the PCIe transfer delay is negligible and unlikely to affect overall task performance.

\section{Related Work}
\myparagraph{VLA Acceleration.}
VLA-Cache~\cite{xu2025vla} accelerate inference by reusing similar vectors from similar inputs.
PD-VLA~\cite{song2025accelerating} uses parallel decode with mathematical properties to accelerate VLA inference.
FAST~\cite{pertsch2025fast} accelerates VLA training by using a new form of action tokenization and transform the training to auto-regressive training.
RoboMamba~\cite{liu2024robomamba} reduces the complexity of model by adapting Mamba~\cite{gu2023mamba} to robotics.
TinyVLA~\cite{wen2025tinyvla} introduces compat vision-language-action model to reduce the calculation size.
QAIL~\cite{park2024quantization} and quantization~\cite{lin2025awq} reduce the robotic model size by quantizing the model weights.
Deer-vla~\cite{yue2024deer} adjusts depth during inference to reduce redundant computation.
Token-pruning~\cite{chen2024image,zhao2024dynamic} prunes the tokens to reduce the inference time.
SparseVLM~\cite{zhang2024sparsevlm} uses sparsity to reduce the computation.
OFT~\cite{kim2025fine} enhances inference efficiency and policy performance by using a new form of fine-tuning.
Actra~\cite{ma2024actra} gives an optimized Transformer architecture to reduce the computation.

\myparagraph{Robotic Generalization.}
Some work~\cite{dey2025redefining,fu2025agentrefine,xiong2024watch} introduce agent to improve the generalization ability of robotic models.
Some work~\cite{shridhar2020alfworld,shinn2023reflexion,dey2025redefining} adds interactive environment to enhance the robot action in unseen states.
Papras~\cite{kim2023papras} designs a plug-and-play system for robot arm system to enhance the ability across different arms.
BC-Z~\cite{jang2022bc} introduces zero-shot task generalization method for robotic imitation learning by algorithm.
Perceiver~\cite{shridhar2023perceiver} designs a multi-task transformer for robotic manipulation.
Copa~\cite{huang2024copa}, Rise~\cite{wang2024rise} and Rh20t~\cite{fang2024rh20t} use spatial constraints to improve generalization.
RST~\cite{li2024robot} designed another data generalization method by self-teaching.
Some work~\cite{ishida2024robust,dosunmu2024demonstrating,li2024experience} optimize the diffusion policy to make it more robust.

\section{Conclusion}
This paper introduces \sys{}, enhancing robotic efficiency and success rate via OS-level primitives. 
It uses \sysexcept{} for quick action interruption, \sysctx{} to avoid redundant computations, 
and \sysreplay{} for effortless repetition. 
Evaluations reveal \sys{} boosts performance by 7-24 $\times$ compared to the robotic system without \sys{} support.